\documentclass[a4paper,twoside,dvips,final,psamsfonts]{amsart}
\usepackage[notref,notcite]{showkeys}
\usepackage{natbib}
\usepackage{amsmath}
\usepackage{amsthm}
\usepackage{epsfig}
\usepackage{amsfonts,amssymb}

\newtheorem{theorem}{Theorem}
\newtheorem{prop}{Proposition}

\theoremstyle{definition}

\theoremstyle{remark}
\newtheorem{remark}{Remark}

\newenvironment{dpmatrix}{\bigl(\smallmatrix}{\endsmallmatrix\bigr)}

\newcommand{\C}{{\mathbb{C}}}

\newcommand{\R}{{\mathbb{R}}}

\newcommand{\one}{\mathbb{I}}
\newcommand{\dom}{{\mathcal{D}}}
\newcommand{\dcar}{{\mathrm{d}}}

\newcommand{\sfrac}[2]{\ensuremath{\frac{{}_{\displaystyle #1}}{{}^{\displaystyle #2}}}}
\newcommand{\ddcar}[2]{\sfrac{\dcar #1}{\dcar #2}}

\DeclareMathOperator{\spec}{spec}
\newcommand{\tr}{\operatorname{tr}}
\newcommand{\tpv}{\operatorname{\mathcal{T}} }
\newcommand{\sgn}{\operatorname{sgn}}

\begin{document}
\title[Spontaneous edge currents for the DIrac equation in 2D]{Spontaneous edge currents for the Dirac equation in two space
dimensions}
\author{Michael J.\ Gruber}
\address[M.~J.~Gruber]{Institut f\"ur Physik, Theoretische Physik II, Universit\"at Augsburg, Germany}

\author{Marianne Leitner}
\address[M.~Leitner]{School of Theoretical Physics, 
Dublin Institute for Advanced Studies, 
Ireland}
\date{Received 20 July 2005, revised 11 November 2005}
\subjclass[2000]{81Q10, 58J32}
\keywords{Dirac operator, boundary condition, Hall effect, spectral flow}

\begin{abstract}
Spontaneous edge currents are known to occur in systems of two space dimensions 
in a strong magnetic field. 
The latter creates chirality and determines the direction of the currents. 
Here we show that an analogous effect occurs in a field-free situation 
when time reversal symmetry is broken by the mass term of the Dirac equation 
in two space dimensions. 
On a half plane, one sees explicitly that the strength of the edge current 
is proportional to the difference between the chemical potentials 
at the edge and in the bulk, 
so that the effect is analogous to the Hall effect, but with an internal potential. 
The edge conductivity differs from the bulk (Hall) conductivity on the whole plane. 
This results from the dependence of the edge conductivity on the choice of a selfadjoint extension of the Dirac Hamiltonian. 
The invariance of the edge conductivity with respect to small perturbations 
is studied in this example by topological techniques.
\end{abstract}

\maketitle

\section{Introduction}

When in a two dimensional device without dissipation, 
an electric field is turned on, 
a current is induced transversally, 
with density subject to the Ohm-Hall law $\vec{j}=\sigma\vec{E}$. 
Here $\sigma$ is the $2\times 2$-conductivity matrix and $\sigma_H:=\sigma_{21}$ defines the Hall conductivity. 
For particles described by a Schr\"odinger operator, 
a magnetic field perpendicular to the plane is needed in addition to obtain $\sigma_H\not=0$ \citep{Asshap:1986}.
However, for more general investigations, 
a time reversal symmetry breaking term in the Hamiltonian 
might suffice to produce a nonzero $\sigma_H$ \citep{Semenoff:1984,Haldane:1988}.
The constant Dirac operator
\begin{equation}\label{Dirac-Op.}
     D= \hbar c(-\imath\vec{\sigma}\cdot\vec{\nabla})+\sigma_3 mc^2
\end{equation}
with fermion mass $m\neq 0$ yields a very instructive example. 
Here $c$ is the velocity of light, 
$\vec\sigma:=(\sigma_1,\sigma_2)$, where $\sigma_i$ are, for $i=1,2,3$, the Pauli matrices, and $\vec\nabla$ is the 2-dimensional gradient.
On $\R^2$, the operator \eqref{Dirac-Op.} features a \emph{zero field Hall effect} \citep{Froehlich:1991}
with $\sigma_H=\tfrac{1}2\sgn(m)\tfrac{e^2}h$ \citep{Redlich:1984}. 
The interpretation of $\sigma_H$ at zero temperature 
as the Chern number of a complex line bundle \citep{TKNN:1982,Kohmoto:1985,AS:1985} fails,
but its quantisation can be traced back to the geometry of the Lorentz group \citep{Lei:ZFHETS,Lei:TBP}.

In the present paper, 
we direct our attention to the Dirac operator \eqref{Dirac-Op.} on a sample with boundary. 
In this situation spontaneous edge currents may occur, without any exterior electric or magnetic field.
We calculate the edge conductivity $\sigma^e$ \citep{Halperin:1982}
for a natural class of self-adjoint extensions of \eqref{Dirac-Op.} on the half-plane.
Here $\sigma^e$ is an integer (in units of $e^2/h$) 
which differs from zero if the boundary condition satisfies a certain sign condition.
It is shown
that $\sigma^e$ is, in units of $e^2/h$, the spectral flow through the gap \citep{Hatsugai1:1993,Hatsugai2:1993}.
Robustness is then immediate for sufficiently small perturbations of \eqref{Dirac-Op.}.\\
In spite of the absence of an exterior field, the edge conductivity can be related
to the Hall conductivity in the bulk. For Schr\"odinger operators in a magnetic field
equality of bulk (Hall) and edge conductivity has been shown in
\citep{KelRicSch:ECCNIQHE,Graf-Elbau:2002}. In our system, the relationship is more subtle,
since the bulk conductivity is half integral, in contrast to the integral edge conductivity.

\begin{figure}[htbp]
\begin{center}
\psfig{figure=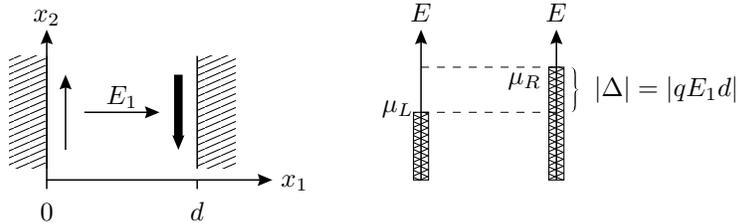}
\end{center}
\caption{
Edge currents in the strip.
The electric field opens an interval $\Delta$ between the respective chemical potentials on the edges. The resulting inequilibrium of charge transport along both edges yields a nonzero total current in the sample.}  
\label{fig:edgemodel}
\end{figure}
Hall currents can go through the bulk or along the edges, 
e.g.\ due to intercepted cyclotron orbits \citep{Schulz-Baldes:2000}.
For $d\gg 1$, consider the strip $[0,d]\times\R\subset\R^2$ with edges of similar type at $x_1=0$ and $x_1=d$, respectively, and no bulk currents. 
When an electric field of constant strength $E_1>0$ is applied, parallel to the $x_1$-axis,
the force $qE_1$ pushes particles away from one edge to the other,
changing the respective chemical potentials 
on the left edge ($\mu_{L}$) and on the right edge ($\mu_{R}$) 
correspondingly (Figure~\ref{fig:edgemodel}). 
For $\mu_R>\mu_L$,
there is a net charge transport  
due to the states with energies contained in the interval $\Delta:=[\mu_{L},\mu_{R}]$,
of width $|\Delta|=|qE_1d|$.
If $j_{\Delta}(x_1)$ denotes the corresponding current density,
the net \emph{edge current} is
\begin{equation}\label{Def. edge current}
J^e_d(\Delta)
:=\int_0^d j_{\Delta}\:dx_1.
\end{equation}
It is related to the voltage $|\Delta|/q$ by
\begin{equation}\label{Def. edge conductivity}
J^e_d(\Delta)
=:\sigma^e(\Delta)\:\frac{|\Delta|}{q}\:.
\end{equation}
Here the proportionality factor $\sigma^e(\Delta)$, given in units of $q^2/h$, defines the \emph{edge conductivity} \citep{Laughlin:1981,Halperin:1982}.
\eqref{Def. edge conductivity} mimicks the Ohm-Hall law $j_2=\sigma^bE_1$ for $\sigma^b:=\sigma_{21}$.\\
For sufficiently large $d$, the two boundaries decouple,
and to calculate $J^e(\Delta)\equiv J^e_{\infty}(\Delta)$,
one only needs to consider a half-plane geometry.
Provided $\mu$ lies in the spectral gap,
with lower gap barrier $E_{\text{crit}}$,
$\Delta:=(E_{\text{crit}},\mu]$.\\ 
Depending on whether the boundary is situated on the left ($x_1=0$) or on the right ($x_1=d$) of the sample, the sign in \eqref{Def. edge conductivity} has to be adjusted, and this is done correctly by imposing $\sgn(\sigma^e(\Delta))=\sgn(\sigma^b)$.
$|\Delta|=\mu-E_{\text{crit}}>0$ can be interpreted as the amount of energy needed 
to excite a bulk particle of energy $E_{\text{crit}}$ to a state at highest possible energy $\mu$. Therefore \eqref{Def. edge conductivity} has the shape of the Ohm-Hall law, 
but here the current is proportional to an interior voltage  (instead of to an exteriorly applied one as in the Hall effect). In particular, $\sigma^e(\Delta)$ is again a conductivity.\\

Our paper is organised as follows: 
In the following section, we introduce the self-adjoint boundary conditions for the constant Dirac operator \eqref{Dirac-Op.} on the half plane. Their effect on the spectrum will be investigated in Section~\ref{sec:spectrum}.
In Section~\ref{sec:edge conductivity}, we derive the corresponding edge conductivity.
Section~\ref{sec:spectral flow} presents a first stability result.

We would like to thank H.~Schulz-Baldes for helpful
discussions.

\section{Boundary conditions}\label{sec:boundary conditions}

As noticed above,
the magnetic field may be zero if a time reversal breaking term in the Hamiltonian is present.
We investigate the Dirac operator \eqref{Dirac-Op.} of massive spin $\frac{1}{2}$ particles (with $q=e$, the electron charge) where this symmetry is broken by the mass term.

$D$ is a symmetric elliptic operator on the domain $\dom(D)=C_c^\infty(\R_+\times\R,\C^2)$ of smooth functions with compact support vanishing in a neighbourhood of $x_1=0$,
but it is not essentially self-adjoint.
Since $D$ is not bounded below the Friedrichs extension is not available for determining a canonical choice of boundary condition.
Note that even in the Schr\"odinger/Pauli case, Dirichlet (Friedrichs) and Neumann boundary condition are not necessarily the boundary condition which represents the physical system best 
\citep[see][where chiral boundary conditions are suggested]{AkkAvrNarSei:BCBESQHS}.
Neither Dirichlet nor Neumann nor chiral provide self-adjoint boundary conditions for Dirac operators.
Therefore, we choose to determine all self-adjoint boundary conditions which respect the symmetry of the problem.

The physical setup is homogeneous w.r.t.\ $x_2$, and so is $D$ on $\dom(D)$.
Fourier transform in $x_2$ gives a unitary transform
\begin{equation}
\begin{aligned}
\Phi: L^2(\R_+\times\R,\C^2) &\rightarrow \int^\oplus_\R L^2(\R_+,\C^2)\,\dcar k_2, \\
(\Phi(\psi))_{k_2}(x_1) &:= \psi_{k_2}(x_1)\quad\text{with}\\
\psi_{k_2}(x_1) &:= \frac1{\sqrt{2\pi}} \int_\R e^{-\imath x_2k_2}\psi(x_1,x_2)\,\dcar x_2.
\end{aligned}\label{eqn:fourierdirect}
\end{equation}
An operator is homogeneous w.r.t.\ $x_2$ if and only if it is decomposable w.r.t.\ the direct integral~\eqref{eqn:fourierdirect} 
\cite[see, e.g.][chapter XIII.16]{ReeSim:AO}.
Of course, we are interested only in those self-adjoint extensions $\tilde D$ of $D$ which preserve homogeneity.
We therefore state
\begin{prop}
The $x_2$-homogeneous self-adjoint extensions $\tilde D$ of $D$ are given exactly by all (measurable) families $\tilde D(k_2)$ of self-adjoint extensions of
$D(k_2)$, where
\begin{equation}
D(k_2) = \sigma_1 \sfrac\hbar\imath c \ddcar{}{x_1} + \sigma_2\hbar c k_2 +\sigma_3mc^2
\label{eqn:diracdirect}
\end{equation}
on $\dom(D(k_2))=C^\infty_c(\R_+,\C^2)$.
\end{prop}
\begin{proof}
Being a differential operator (with smooth coefficients), $D$ is a closable operator.
By continuity the closure $\bar D$ is homogeneous, and for closed operators we have the equivalence between homogeneity and decomposability cited above.
The fibres $\bar D(k_2)$ of $\bar D$ are closed, and $C^\infty_c(\R_+,\C^2)$ is clearly an operator core for $\bar D(k_2)$.
This proves the first part.

The second part is a standard calculation with the Fourier transform.
\end{proof}

For determining the self-adjoint extensions of $D(k_2)$ for fixed $k_2$ we follow the von Neumann theory of extensions
\cite[see, e.g.,][chapter X.1]{ReeSim:FASA}:
\begin{theorem}
\label{theorem:saextensions}
The self-adjoint extensions of $D(k_2)$ are parametrized by $\zeta\in \overline\R:=\R\cup\{\infty\}$.
The extension $D_{\zeta}(k_2)$ is given by the domain
\begin{equation}
\dom(D_{\zeta}(k_2)) = \left\{ \begin{pmatrix}
v\\w
\end{pmatrix} \in H^1(\R_+) \colon w(0)=\imath\zeta v(0) \right\}
\label{eqn:cdomain}
\end{equation}
where $\zeta=\infty$ is understood to mean $v(0)=0$, and $H^1$ denotes the $L^2$-Sobolev space of order 1.
\end{theorem}
Note that, by Sobolev's embedding lemma, $H^1$-functions on $\R_+$ are continuous, so that $v(0)$ makes sense.
Physically,
\eqref{eqn:cdomain} says that at $x_1=0$, 
no current perpendicular to the boundary is allowed. 
Indeed, $j_1=ev_1$ with the velocity operator $v_1:=\frac{1}{\imath\hbar}[x_1,H]=c\sigma_1$ 
acting on $\C^2$. Now the matrix element
\begin{equation*}
\begin{pmatrix}
v & w
\end{pmatrix}
\sigma_1
\begin{pmatrix}
v \\ w
\end{pmatrix} = \bar v w + \bar w v = 2\Re(\bar v w)
\end{equation*}
vanishes if and only if $w=\imath\zeta v$ for $\zeta\in \overline\R$.

\begin{proof}
The bounded parts do not matter for questions of self-adjointness (they do change the parametrization)
and we choose units with $\hbar=1,c=1$ for this proof
so that we have to deal with $T:=D(k_2)=\sigma_j\sfrac1\imath\ddcar{}x$ only ($j=1$, $x=x_1$).

Since $T$ is first order differential and elliptic, the adjoint is given by the domain $\dom(T^*)=W^1(\R_+)$ (i.e.\ no boundary conditions).
According to von Neumann's theorem we have to compute the $\pm\imath$ eigenspaces of $T^*$.
Because of ellipticity they are given by smooth functions, because of uniqueness they are at most one-dimensional.
We have
\[ T^*\psi=\pm\imath\psi
\Leftrightarrow \psi'=\mp \sigma_j \psi \Rightarrow \psi''=\psi \]
so that $\psi(x) = \begin{dpmatrix} a \\ b \end{dpmatrix} e^{-x} $ for some constants $a,b\in\C$.
Reinserting this into the eigenvalue equation yields
\begin{equation}
\sigma_j \begin{pmatrix} a \\ b
\end{pmatrix} = \pm \begin{pmatrix} a \\ b \end{pmatrix}
\label{eqn:sigmaeigen}
\end{equation}
which is an easily solvable eigenvalue problem in $\C^2$.
$P_j^\pm:=\tfrac12(1\pm\sigma_j)$ are the corresponding eigenprojections.
To sum up, the $\pm\imath$ eigenspaces of $T^*$ are given by
\[ K^\pm = P_j^\pm\C^2\, e^{-x}. \]

Now we have to find all unitaries $K^+\to K^-$.
Since $K^\pm$ are one-dimensional, all unitaries differ only by a complex number $z$ of modulus $1$.
If $k\ne j$ then $\sigma_k\sigma_j=-\sigma_j\sigma_k$ by the canonical anti-commutation relations for Pauli matrices.
So, $\sigma_k P_j^\pm = P_j^\mp \sigma_k$.
Therefore, $\sigma_k$ maps $K^+$ to $K^-$ and vice versa, and it is clearly a unitary, so that all unitaries are of the form $U_z=z\sigma_k$.

Again, according to von Neumann theory, to each $U_z$ corresponds a self-adjoint extension $T_z$ with domain
\begin{align}
\dom(T_z) &= \dom(\bar T) \oplus \left\{ (1-U_z)\psi \colon \psi\in K^+ \right\} \\
 &=  \dom(\bar T) \oplus \left\{ (1-z\sigma_k) \begin{pmatrix} a \\ b \end{pmatrix} e^{-x} \colon \sigma_j \begin{pmatrix} a \\ b \end{pmatrix} = \begin{pmatrix} a \\ b \end{pmatrix} \right\}.
\end{align}
Note that
\[ \sigma_j \begin{dpmatrix} a \\ b \end{dpmatrix} = \begin{dpmatrix} a \\ b \end{dpmatrix} \Leftrightarrow P_j^- \begin{dpmatrix} a \\ b \end{dpmatrix} = 0
\Leftrightarrow P_j^+ \begin{dpmatrix} a \\ b \end{dpmatrix} = \begin{dpmatrix} a \\ b \end{dpmatrix} \]
so that
\[
\psi\in\dom(T_z) \Leftrightarrow \psi(0)= (1-z\sigma_k) \begin{dpmatrix} a \\ b \end{dpmatrix}\text{ and } P_j^-\begin{dpmatrix} a \\ b \end{dpmatrix} =0
\]
(and $\psi\in H^1$, of course).
In other words, the possible boundary values $\psi(0)$ are given by the range of $R:=(1-z\sigma_k)P_j^+$ which is a non-othogonal projection.
Furthermore,
\[ P_j^- (1+z\sigma_k) = \tfrac12 (1-\sigma_j)(1+z\sigma_k) = 1 - (1-z\sigma_k)P_j^+ \]
so that the self-adjoint boundary condition can be equivalently described by noting
\begin{equation}
P_j^- (1+z\sigma_k) \psi(0) = 0 \Leftrightarrow \psi(0)=(1-z\sigma_k)P_j^+ \psi(0)
\label{eqn:boundarycondition}
\end{equation}
which we will use in Section~\ref{sec:spectrum}.

For $j=1$ and, say, $k=3$, one computes easily $R\begin{dpmatrix} 1 \\ 0 \end{dpmatrix}=\tfrac12 \begin{dpmatrix} 1+z \\ 1-z \end{dpmatrix}$ which is nonvanishing so that it spans the one-dimensional space of boundary values $\psi(0)=\begin{dpmatrix} v \\ w \end{dpmatrix}$.
So we arrived at
\[ w = \frac{1+z}{1-z} v \]
which is a fractional linear transformation in $z$, and as such maps circles to lines or circles.
Inserting a few values on the circle $|z|=1$ one sees that it is mapped indeed to the line $\imath\zeta$, $\zeta\in\R$.
\end{proof}

Note that, in principle, the parameter $\zeta$ specifying the boundary condition is allowed to vary with $k_2$ without breaking homogeneity.
In the following we restrict ourselves to constant $\zeta$, even though the discussion of the spectrum (except for the pictures) goes through in the general case as well.

\section{Spectrum}\label{sec:spectrum}
Note that $D_{\zeta}(k_2)$ depends continuously on $k_2$ so that, by the standard theory of direct integrals, the spectrum of $D_{\zeta}$ is given by
\begin{equation}
\spec D_{\zeta}= \bigcup_{k_2\in\R} \spec D_{\zeta}(k_2).
\label{eqn:specdirect}
\end{equation}
The spectrum of the fibre operator $D_{\zeta}(k_2)$ is determined in the following:
\begin{theorem}
\label{theorem:spectrum}
The spectrum of $D_{\zeta}(k_2)$ consists of:
\begin{enumerate}
\item a continuous part $\{E \colon E^2\geq E_b(k_2)^2\}$, 
where $E_b=\sqrt{(\hbar ck_2)^2+(mc^2)^2}$
(bulk part) and \label{enum:continuous part}
\item a gap eigenvalue $\displaystyle E_g(k_2)=\frac{2\zeta\hbar c k_2+(1-\zeta^2)mc^2}{1+\zeta^2}$ under the condition \label{enum:gap eigenvalue}
\begin{equation}
\hbar k_2(\zeta^2-1)> -2mc\zeta. \label{eqn:gapcondition}
\end{equation}
\end{enumerate}
\end{theorem}
\begin{proof}
Again we choose the simplified notation from the proof of Theorem~\ref{theorem:saextensions}
and write $T=D_{\zeta}(k_2)$.
If $E$ is an eigenvalue of $T$ then $E^2$ is an eigenvalue of
\begin{equation}
 T^2 = -\ddcar{ {}^2}{x^2} +k_2^2+m^2 \label{eqn:t2eigen}
\end{equation}

We begin with the case $E^2<k_2^2+m^2$. The only bounded solutions $\psi$ of $T^2\psi=E\psi$ have the form
\begin{equation}
\psi(x) = \begin{pmatrix} a\\b \end{pmatrix} e^{-x\sqrt{k_2^2+m^2-E^2}}
\label{eqn:eigensolution}
\end{equation}
with arbitrary $a,b\in\C$.
Plugging this into the eigenvalue equation $T\psi=E\psi$ gives the condition
\begin{align}
 Q_E  \begin{pmatrix} a\\b \end{pmatrix} &= E  \begin{pmatrix} a\\b \end{pmatrix}\text{ with} \\
 Q_E &= \imath\sqrt{k_2^2+m^2-E^2}\sigma_1 +k_2\sigma_2+m\sigma_3
\end{align}
in addition to the boundary condition.
Note that
\[ Q_E^2= -(k_2^2+m^2-E^2)+k_2^2+m^2=E^2 \]
 and $\tr Q_E=0$ so that the matrix $Q_E$ has spectrum $\{\pm E\}$ and there is always a nontrivial solution.
For $E\ne0$ we define a corresponding (non-orthogonal) eigenprojection $P_E:=\tfrac12(1+\tfrac1EQ_E)$ (the case $E=0$ is dealt with easily).
All candidates for eigensolutions are within the range of $P_E$.
On the other hand, the boundary condition in the form \eqref{eqn:boundarycondition} requires $P_1^-(1+z\sigma_3)\psi(0)=0$.
A straightforward computation with Pauli matrices results in
\begin{align}
A:=P_1^-(1+z\sigma_3)P_E &= \frac1{4E}(v-v\sigma_1+w\sigma_2-\imath w\sigma_3) \text{ where} \\
  v &= E-\imath\sqrt{k_2^2+m^2-E^2} + zk_2\imath+zm, \\
  w &= k_2+zE\imath -z\sqrt{k_2^2+m^2-E^2}+\imath m.
\end{align}
The condition for the existence of a nontrivial eigensolution fulfilling the boundary condition is therefore $A=0$, since $P_E$ has one-dimensional range onl
y.
Closer inspection shows $w=\imath z \bar v$ so that $v=0$ is the only condition to check.
(Note that the Pauli matrices form a basis of $M(2,\C)$.)
\begin{align}
v=0 &\Leftrightarrow E+zk_2\imath+zm = \imath\sqrt{k_2^2+m^2-E^2} \\
  &\Leftrightarrow \Re(E+zk_2\imath+zm)=0 \text{ and } \Im(E+zk_2\imath+zm)\geq 0
\end{align}
 From this we get
\begin{equation}
E=k_2 \Im z - m \Re z = \frac{2\zeta k_2+m(1-\zeta^2)}{1+\zeta^2}
\end{equation}
and
\begin{equation}
0 \leq \Im(zk_2\imath+zm)=k_2\Re z+m\Im z = \frac{k_2(\zeta^2-1)+2m\zeta}{1+\zeta^2}
\end{equation}
which proves the claim about the gap spectrum.

In the case $E^2>k_2^2+m^2$ there are always two bounded solutions
$\psi_\pm$ of $T^2\psi=E\psi$, having the form
\begin{equation}
\psi_\pm(x) = \begin{pmatrix} a_\pm\\b_\pm \end{pmatrix} e^{\pm\imath x\sqrt{E^2-k_2^2-m^2}}
\label{eqn:eigensolutionb}
\end{equation}
with arbitrary $a_\pm, b_\pm\in\C$,
so that we have to define two matrices $Q_{E,\pm}$ and two corresponding projections $P_{E,\pm}$.
Together with the boundary condition this gives the requirement
\[ 0=P_1^-(1+z\sigma_3)\left( P_{E,+} \begin{pmatrix} a_+\\b_+ \end{pmatrix} + P_{E,-} \begin{pmatrix} a_-\\b_- \end{pmatrix}  \right) \]
which has always nontrivial solutions since this is a linear map $\C^4\to\C^2$.
This proves the claim about the bulk spectrum.
\end{proof}

\begin{remark}
For the system on $\R^2$, $D(k_2)$ lives on $\R$, and its spectrum consists of
$\{E \colon E^2\geq E_b(k_2)^2\}$ only since the solutions for other energies increase exponentially either at $x=\infty$ or $x=-\infty$.
This explains the term bulk spectrum because $\R^2$ is the configuration space of a bulk system.

For fixed $k_2$ the bulk spectrum of our $D_{\zeta}$ has a gap $(-E_b(k_2),E_b(k_2))$.
According to \eqref{eqn:specdirect}, it is $\Delta:=(-|m|c^2,|m|c^2)$.
This is the gap we will be interested in.
\end{remark}

\begin{figure}[htbp]
\begin{center}
\psfig{figure=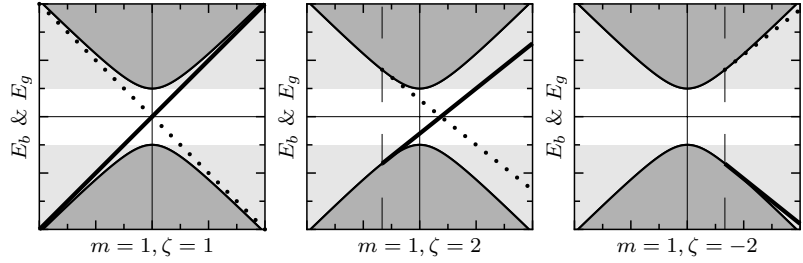}
\end{center}
\caption{Spectrum of $D_{\zeta}(k_2)$ for different $m,\zeta$. The thick lines are $E_g$ for $m,\zeta$ as indicated, the dotted lines are $E_g$ for $-m,-\zeta$. The dashed line indicates $k_{crit}$ (see Proposition~\ref{Prop: mc condition}).}
\label{fig:spectrum}
\end{figure}

\begin{prop}\label{Prop: mc condition}
As $k_2$ varies over $(-\infty,\infty)$, the gap eigenvalue $E_g(k_2)$ goes through the gap $(-|m|c^2,|m|c^2)$ if and only if $m\zeta>0$, i.e. when $\sgn m = \sgn\zeta$.
\end{prop}
\begin{proof}
If $\zeta^2=1$ then the gap condition~\eqref{eqn:gapcondition} requires $m\zeta\geq0$, and $E_g(k_2)=\zeta\hbar k_2c$.
This gives $m\zeta>0$.

If $\zeta^2>1$ then the gap condition requires $k_2\geq k_{crit}$ with $k_{crit}:=-\frac{2mc\zeta}{\hbar(\zeta^2-1)}$.
Note that $k_{crit}$ is exactly the value of $k_2$ where the line $E_g(k_2)$ hits the hyperbola $E_b(k_2)$.
Therefore, $E_g$ goes through the gap if and only if $k_{crit}<0$, which is equivalent to $m\zeta>0$.

If $\zeta^2<1$ then the gap condition requires $k_2\leq k_{crit}$.
Therefore, $E_g$ goes through the gap if and only if $k_{crit}>0$, which is equivalent to $m\zeta>0$ again (note that $\zeta^2-1<0$ in the present case, so that the
direction of the inequality changes again).
\end{proof}

\section{Edge conductivity on the half plane}\label{sec:edge conductivity}

For the constant Dirac operator \eqref{Dirac-Op.} over $\R^2$, 
the bulk conductivity is \citep{Redlich:1984,Ludwig:1994,Lei:ZFHETS,Lei:TBP}
\begin{equation}\label{bulk conductivity}
\sigma^b
=\frac{1}{2}\sgn(m) 
\end{equation}
in units of $e^2/h$. 
To study the corresponding edge conductivity on the half plane, 
for $\zeta\in\R$, let $D_{\zeta}\equiv\{D_{\zeta}(k_2)\}_{k_2\in\R}$ be the operator family defined by Theorem~\ref{theorem:saextensions}.

\begin{theorem}\label{theorem: edge conductivity}
Let $\Delta$ be the gap of the bulk spectrum of $D_{\zeta}$.
Then for any nonempty subinterval $\Delta'\subseteq\Delta$,
the edge conductivity $\sigma^e(\Delta')$ is,
in units of $\frac{e^2}{h}$,
\begin{equation}\label{Gl: result for sigma}
\sigma^e(\Delta')
=\begin{cases} \sgn(m) & \text{if }m\zeta>0, \\
                    0 & \text{otherwise}.
 \end{cases}
\end{equation}
In particular, $\sigma^e(\Delta')\equiv\sigma^e$ does not depend on the choice of $\Delta'\subseteq\Delta$.
$\sigma^e$ is the spectral flow through $E=0$ of $D_{\zeta}$.
\end{theorem}
\begin{remark}
The edge conductivity on the half-plane 
equals the bulk conductivity \eqref{bulk conductivity} on $\R^2$ 
in the sense that $\sigma^b$ 
is the arithmetic mean value of the two possible values for $\sigma^e$.
\end{remark}

Note that interchanging the r\^ oles of $x_1$ and $x_2$ amounts to rotating the sample by $\pi/2$ and to multiplying $\zeta\in\R$ by $\imath$ in the complex plane. 
If $\zeta\not=0$, this yields a proportionality factor $\tilde{\zeta}\in\R$ of sign $-\sgn(\zeta)$,
and, in terms of of $\tilde{\zeta}$, 
the inequality in the gap condition of Proposition \ref{Prop: mc condition} is reversed. However, this modification leaves $\sigma^e$ unaffected because of the sign convention used in \eqref{Def. edge conductivity}.
\begin{proof}
We will proceed in two ways. First, 
let $\psi_{k_2}(x_1)$ be the normalised eigenfunctions \eqref{eqn:eigensolution} of $D_{\zeta}(k_2)$.
Eq.\ \eqref{Def. edge current} yields
\begin{equation}\label{eqn: edge current in Fourierspace}
J^e(\Delta)
=ec\int_{\{k_2:E(k_2)\in\Delta\}}
\langle\psi_{k_2}\mid\sigma_2\mid\psi_{k_2}\rangle_{L^2(\R_+)}\:\frac{dk_2}{2\pi}.
\end{equation}
Using $v_2(k_2)=\hbar^{-1}dD_{\zeta}(k_2)/dk_2$ and the normalisation condition, we obtain
\begin{equation}\label{slope}
\langle\psi_{k_2}\mid\sigma_2\:\psi_{k_2}\rangle_{L^2(\R_+)}
=\frac{1}{c\hbar}\frac{dE_g(k_2)}{dk_2}=\frac{2\zeta}{\zeta^2+1}
\end{equation}
from Theorem \ref{theorem:spectrum}.
\eqref{slope} shows that $\langle\psi_{k_2}\mid j_2\mid\psi_{k_2}\rangle_{L^2(\R_+)}$ does not depend on $k_2$,
so that by \eqref{Def. edge conductivity},
\begin{equation}\label{absolute value of the inverse of the slope}
\frac{h}{e^2}\:\sigma^e(\Delta)
\:\propto\:\frac{c}{\hbar}\:|\Delta|^{-1}\int_{E(k_2)\in\Delta}dk_2
\end{equation}
with proportionality factor \eqref{slope}. 
But r.h.s.\ of \eqref{absolute value of the inverse of the slope} is just the absolute value of the inverse of the slope of the line $E_g(k_2)$. 
Taking Proposition \ref{Prop: mc condition} into account, we conclude (\ref{Gl: result for sigma}). For the last statement,
rewrite \eqref{Def. edge current} as 
\begin{equation*}
J^e(\Delta) =  e\tau(\one_\Delta(H)v_2),
\end{equation*}
where $v_2=\frac{1}{\imath\hbar}[x_2,D]=:\frac 1\hbar\partial_2 D$ and $\one_\Delta(D)$ being the spectral projection of $D$ onto $\Delta$.
$\tpv_2$ is the trace per unit volume in direction $x_2$ for homogeneous operators $A$, 
defined as
\begin{equation}
\tpv_2(A) = \frac1{2\pi}\int_\R A(k_2)\,\dcar k_2,
\label{eqn:tpv}
\end{equation}
where $\int^\oplus_{\R} A(k_2)\dcar k_2=\Phi A\Phi^{-1}$, and $\tr_1$ is the ordinary trace in direction $x_1$ (including the spin-trace over $\C^2)$.
Now approximate $\one_\Delta/|\Delta|$ by $g'$ for a \textit{switch function} $g\in C^\infty(\overline{\underline{\R}})$ (denote $\overline{\underline{\R}}:=\R\cup\{\pm\infty\}$) with $g'\geq 0$, $\operatorname{supp} g'\subset\Delta$, $g(\infty)=1$, $g(-\infty)=0$ \cite[see, e.g.,][]{KelRicSch:ECCNIQHE}.
Then
\begin{equation*}
J^e(\Delta) = e \tau\left(\one_\Delta(D)v_2  \right) \approx e |\Delta | \tau\left(g'(D)v_2  \right) = |\Delta|\frac{e}{\imath\hbar} \tau\left(g'(D)\partial_2D  \right).
\end{equation*}
Denote by $\psi_{k_2}$ a normalised eigenvector for $E_g(k_2)$, differentiable in $k_2$. Then
    \begin{align*}
   \tr_1(g'(D)\partial_2D)(k_2) & = g'(E_g(k_2)) \langle \psi_{k_2}|\partial_{k_2} D(k_2)\psi_{k_2}\rangle_{L^2(\R_+)} \\
 &= g'(E_g(k_2)) \ddcar{}{k_2} E_g(k_2) = \ddcar{}{k_2} g(E(k_2)),
    \end{align*}
whose integral is $\sigma^e(\Delta)$ as given by (\ref{Gl: result for sigma}), in units of $\frac{e^2}{h}$. This proof also shows the topological nature of the result.\\
\end{proof}

The essential point is that instead of varying the subinterval $\Delta'\subseteq\Delta$ but using the eigenvalue dispersion explicitely, 
the second approach keeps the calculation quite general by introducing a function $g$ which we allow to vary (while now the gap interval is fixed).
The topological nature of $\sigma^e$ will enable us to show the invariance of $\sigma^e$ at least under a simple class of perturbations.

\section{Spectral flow and stability}\label{sec:perturbations}\label{sec:spectral flow}
One of the most remarkable properties of the integer QHE is its stability w.r.t.\ perturbations (disorder). The simplest case is when the perturbation depends on $x_1$ only:
\begin{prop}\label{prop:perturbation}
Let $W$ be a bounded self-adjoint operator on $L^2(\R_+,\C^2)$, inducing a homogeneous (w.r.t.\ $x_2$) bounded operator on $L^2(\R_+\times\R,C)$.
If $\|W\|<|m|$  then the system described by $D_{\zeta}+W$ has the same edge conductivity
as one described by $D_{\zeta}$.
\end{prop}
\begin{proof}
First note that $W$, being  bounded, does not change anything regarding the boundary conditions and self-adjoint extensions.
Since $W$ is independent of $x_2$, the direct integral decomposition of $D_{\zeta}+W$ is $D_{\zeta}(k_2)+W$, and therefore the Hall conductivity is given by the spectral
flow as before.

Through addition of $W$, the spectrum of $D_{\zeta}k_2)$ can change by $\pm\|W\|$ only.
Therefore a gap around $0$ in the bulk spectrum remains as long as $\|W\|<|m|$.
In the same way, in the $\|W\|$-neighbourhood of $E_g(k_2)$ there will be a unique eigenvalue of $D_{\zeta}(k_2)+W$ if $m\zeta>0$.
Since $E_g(k_2)$ goes from below $-|m|$ to above $|m|$ or vice versa, the unique eigenvalue in the perturbed system will cross $0$ in the same direction as long as $\|W\|<|m|$.
Thus the spectral flow is the same.
\end{proof}
Note that $W$ is not restricted to be multiplication by a function.
Choosing $W=m_1(x_1)\sigma_3+V(x_1)$ with bounded (smooth, for simplicity) $m_1,V$ allows for variable mass and electric potential.

We now turn to the more general case of perturbations which are periodic in $x_2$.
Since $D$ is not homogeneous w.r.t.\ $x_2$ any more, we have to replace
Fourier transform w.r.t.\ $x_2$ as in \eqref{eqn:fourierdirect} by Floquet-Bloch analysis w.r.t.\ $x_2$ \citep[see, e.g.,][chapter XIII.16]{ReeSim:AO}.
Then, for a periodic operator $A$ on $L^2(\R_+\times\R,\C^2)$, its
Floquet-Bloch transform $A(k_2)$  acts on $L^2(\R_+\times [0,L],\C^2)$ with $k_2$-quasiperiodic boundary conditions on $[0,L]$, 
and the trace per unit volume is
\begin{equation}
\tpv_2(A) = \frac1{2\pi}\int_{[-\pi/L,\pi/L]} \tr_{L^2[0,L]} A(k_2)\,\dcar k_2.
\label{eqn:tppv}
\end{equation}
Note that homogeneous operators are in particular periodic, and that for these, Definition \eqref{eqn:tppv} gives the same trace as \eqref{eqn:tpv}. 
\begin{figure}[htbp]
\begin{center}
\psfig{figure=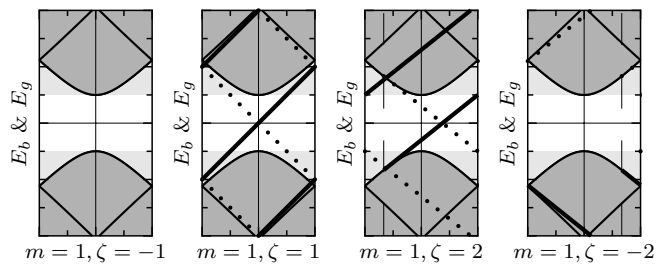}
\end{center}
\caption{Spectrum in the first Brillouin zone $[-\pi/L,\pi/L]$. Dashed and dotted lines have the same meaning as in Figure~\ref{fig:spectrum}.}
\label{fig:periodic spectrum}
\end{figure}

Reviewing the spectral results from Section~\ref{sec:spectrum} in the framework of the Bloch-Floquet decomposition leads to the spectrum shown in Figure~\ref{fig:periodic spectrum}. 
Note how in this representation (so called reduced zone scheme) the bands and eigenvalues are mapped back periodically to the $k_2$-interval $[-\pi/L,\pi/L]$.

Now, going through the arguments above we see that $\sigma^e$
is still given by the spectral flow, even when computed through the Bloch-Floquet decomposition.
Therefore, Proposition~\ref{prop:perturbation} holds mutatis mutandis.

For physical applications one would like stability under random perturbations describing disorder in a crystal.
If $W$ is random we cannot apply the Bloch-Floquet decomposition any more.
Instead, one could use techniques from Non-Commutative Geometry as was done in \citep{BelElsSch:NGQHE} for the quantum Hall-effect.
It would be interesting to allow randomness in the boundary condition $\zeta$ as well since this would describe surface imperfections.
We leave this to future work.




\ifx\undefined\allcaps\def\allcaps#1{#1}\fi

\end{document}